\documentclass[reprint,aps,pra,showkeys,unsortedaddress,superscriptaddress]{revtex4-1}
\usepackage{subfiles}
\usepackage{graphicx}
\usepackage{amsfonts,amsmath}
\graphicspath{{Figures/Figures/}}
\usepackage[margin = 0.5in]{geometry}
\usepackage{romannum}

\begin{document}
\title{Direct quantification of quasi-Fermi level splitting in organic semiconductor devices}
\author{Drew B. Riley}
\email[Email Drew B. Riley at: ]{1915821@swansea.ac.uk}
\author{Oskar J. Sandberg}
\email[Email Dr. Oskar J. Sandberg at: ]{o.j.sandberg@swansea.ac.uk}
\affiliation{Sustainable Advanced Materials Programme (Sêr SAM), Department of Physics, Swansea University, Singleton Park, Swansea SA2 8PP, United Kingdom}
\author{Nora M. Wilson}
\affiliation{Faculty of Science and Engineering, \AA bo Akademi University, 20500 Turku, Finland}
\author{Wei Li}
\author{Stefan Zeiske}
\author{Nasim Zarrabi}
\author{Paul Meredith}
\affiliation{Sustainable Advanced Materials Programme (Sêr SAM), Department of Physics, Swansea University, Singleton Park, Swansea SA2 8PP, United Kingdom}
\author{Ronald \"Osterbacka}
\affiliation{Faculty of Science and Engineering, \AA bo Akademi University, 20500 Turku, Finland}
\author{Ardalan Armin}
\email[Email Dr. Ardalan Armin at: ]{ardalan.armin@swansea.ac.uk}
\affiliation{Sustainable Advanced Materials Programme (Sêr SAM), Department of Physics, Swansea University, Singleton Park, Swansea SA2 8PP, United Kingdom}

\begin{abstract}
Non-radiative losses to the open-circuit voltage are a primary factor in limiting the power conversion efficiency of organic photovoltaic devices. The dominant non-radiative loss is intrinsic to the active layer and can be determined from the quasi-Fermi level splitting (QFLS) and the radiative thermodynamic limit of the photovoltage. Quantification of the QFLS in thin film devices with low mobility is challenging due to the excitonic nature of photoexcitation and additional sources of non-radiative loss associated with the device structure. This work outlines an experimental approach based on electro-modulated photoluminescence, which can be used to directly measure the intrinsic non-radiative loss to the open-circuit voltage; thereby, quantifying the QFLS. Drift-diffusion simulations are carried out to show that this method accurately predicts the QFLS in the bulk of the device regardless of device-related non-radiative losses. State-of-the-art PM6:Y6-based organic solar cells are used as a model to test the experimental approach, and the QFLS is quantified and shown to be independent of device architecture. This work provides a method to quantify the QFLS of organic solar cells under operational conditions, fully characterizing the different contributions to the non-radiative losses of the open-circuit voltage. The reported method will be useful in not only characterizing and understanding losses in organic solar cells, but also other device platforms such as light-emitting diodes and photodetectors.

\end{abstract}
\maketitle

\section{Introduction}
As power conversion efficiencies of organic solar cells surpass 18\% \cite{lin2020,liu2020-2} it has become essential to comprehend and eradicate every mechanism contributing to efficiency reduction. Losses to the short-circuit current ($J_\text{sc}$) have been minimized by increasing charge generation and collection through the use of non-fullerene acceptors, resulting in short circuit-currents consistently above 20 mA/$\text{cm}^2$ \cite{cui2019,liu2020-2,perdigon2020,armin2021}. While open-circuit voltages ($V_\text{oc}$) have also increased in non-fullerene acceptors, the mechanisms contributing to $V_\text{oc}$ losses are not as straightforward to interpret as those contributing to the short-circuit current losses. As losses in the $V_\text{oc}$ due to radiative recombination channels are unavoidable in solar cells, the radiative limit to the $V_\text{oc}$ ($V_\text{oc}^\text{Rad}$) is considered the primary benchmark to target. Non-radiative losses to the $V_\text{oc}$ result in proportional decreases to the power conversion efficiency, making their detection and suppression a high research priority \cite{vandewal2008,vandewal2009,vandewal2010,vandewal2012,elumalai2016,baran2016,scharber2006, mihailetchi2003}. Non-radiative recombination losses can be intrinsic to the device's active layer due to the interaction of electronic states with the surrounding medium via vibrational states. 
Additionally, non-radiative losses can occur at the interfaces between the active layer and the electrode (or interlayer). This non-radiative recombination channel, defined by the extraction of minority carriers out of the active layer at the `wrong' contact (i.e. electrons at the anode, holes at the cathode), is often referred to as surface recombination \cite{wagenpfahl2010,knesting2013,kirchartz2013,reinhardt2014,hacene2014,wheeler2015,zonno2016,sandberg2019,karki2020}. 
As the $V_\text{oc}$ is ultimately defined by the quasi-Fermi level splitting (QFLS) of electrons and holes in the device (at open-circuit), an accurate quantification of the QFLS is key for understanding non-radiative recombination processes in photovoltaic devices.

Quasi-Fermi levels and QFLS, as introduced by Shockley, is an essential concept used to describe non-equilibrium steady-state operation of electronic semiconductors \cite{shockley1949}. This concept was expanded by W$\ddot{\text{u}}$rfel for opto-electronic processes by including a chemical potential associated with the radiative process \cite{wurfel1982}. W$\ddot{\text{u}}$rfel showed that the chemical potential of an absorbed or emitted photon ($\mu_\gamma$) is equal to the QFLS in the semiconductor. Equipped with this understanding researchers have successfully employed photoluminescence measurements to evaluate the QFLS in systems where absorption and emission are dominated by free carriers \cite{wurfel1982,braly2018,stolterfoht2019}, and subsequently applied this technique to organic semiconductors \cite{cao1999,mitschke2000,grimsdale2009}. Regrettably, photoexcitation in organic semiconductors is not dominated by free carriers but by Coulombically bound singlet excitonic states \cite{bakulin2012}. To generate free carriers excitons must diffuse to a donor-acceptor interface and form intermediary charge-transfer (CT) states, which primarily decay non-radiatively \cite{classen2020}. While CT states may be in equilibrium with free carriers, excitons are generally not and may decay radiatively before forming a CT state \cite{vandewal2008,vandewal2009,vandewal2010,vandewal2012,perdigon2020}. Therefore, traditional photoluminescence measurements (as applied to inorganic semiconductors) are not valid as a method to determine the QFLS in organic semiconductor blends since the contribution to charge generation is simultaneously over-estimated for excitons and under-estimated for CT states. These discrepancies can be circumvented by employing Rau's reciprocity principle between the charge collection of photogenerated carriers (under illumination) and the electroluminescent emission in the dark. Rau's theory assesses the non-radiative losses by providing an expression for both the radiative limit and the non-radiative losses of the open-circuit voltage of a device \cite{rau2007}. This has been successfully employed in conjunction with electroluminescent external quantum efficiency measurements to quantify the QFLS and related losses in a wide variety of semiconductor-based solar cells  \cite{green2001,kirchartz2007,kirchartz2008,kirchartz2009}. However, the corresponding $V_\text{oc}$ loss derived from this approach generally includes contributions from both intrinsic bulk-related processes and surface recombination. 
In organic solar cells, the presence of surface recombination is usually correlated with electrode-induced photovoltage losses at the contacts \cite{sandberg2016,spies2017,phuong2020}, causing a mismatch between $V_\text{oc}$ and the associated QFLS in the bulk. As such, measurements based on electroluminscence cannot differentiate between intrinsic and electrode-induced photovoltage losses, leading to a consistent underestimation the QFLS in organic photovoltaic devices. A method for overcoming the difficulty in quantifying QFLS has been recently suggested using photo-induced absorption \cite{phuong2020}. However, this method requires detailed knowledge about the absorption cross-section and charge transport parameters of the device, which must necessarily be semi-transparent. A more direct quantification of the QFLS in an optimized device would serve well to complement this method and provide quantification of QFLS in devices where photo-induced absorption is not trivial.

The work described herein establishes a novel experimental approach labelled electro-modulated photoluminescence quantum efficiency, providing a pathway to measuring the intrinsic non-radiative losses occurring within the active layer of an organic solar cell at open-circuit conditions. Through the principle of reciprocity this technique quantifies the QFLS in the active layer and subsequently the electrode-induced voltage losses of a solar cell under operational conditions. Drift-diffusion simulations are employed to simulate the proposed experiment and compare to the computable QFLS. Further, PM6:Y6 (see S \Romannum{3} for a list of chemical acronyms) solar cells have been fabricated with various contacts designed to increase the non-radiative loss without modifying the QFLS in the bulk to confirm the simulated experiments. The electro-modulated photoluminescence measurements were found to successfully predict the QFLS and related losses over the range of devices used. 

\section{Theoretical Background}
For a semiconductor with flat featureless quasi-Fermi levels, the emission flux ($\Phi_\text{em}$) of photons with energy $E_\gamma$ is determined by the chemical potential of radiation $\mu_\gamma$ via \cite{wurfel1982}:
\begin{equation}\label{equation:Wurfel}
\Phi_\text{em}\left(E_\gamma\right) = a\left(E_\gamma\right)\Phi_\text{bb}\left(E_\gamma\right)\left[\exp\left(\frac{\mu_\gamma}{k_\text{B}T}\right)-1\right]
\end{equation}
assuming $E_\text{G} - \mu_\gamma \gg k_\text{B}T$, where $E_\text{G}$ is the bandgap of the semiconductor, $k_\text{B}$ is the Boltzmann constant, $T$ is the temperature of the lattice, $a\left(E_\gamma\right)$ is the spectral absorbance, and $\Phi_\text{bb}\left(E_\gamma\right)$ is the spectral black-body radiation at room temperature. Using the reciprocity principle an analogous relation between the emitted radiation, photovoltaic external quantum efficiency ($\text{EQE}_\text{PV}$), and voltage ($V$) can be expressed as \cite{rau2007}:
\begin{equation}\label{equation:Rau}
\Phi_\text{em}\left(V,E_\gamma\right) = \text{EQE}_\text{PV}\left(E_\gamma\right)\Phi_\text{bb}\left(E_\gamma\right)\left[\exp\left(\frac{qV}{k_\text{B}T}\right)-1\right]
\end{equation}
where $q$ is the elementary charge. Note that the substitution of $\text{EQE}_\text{PV}$ for $a$ and $qV$ for $\mu_\gamma$ in Eq.~\ref{equation:Rau} produces Eq.~\ref{equation:Wurfel}. However, this substitution is only valid when the charge carriers in the semiconductor are in equilibrium with the emitted radiation across the device structure, satisfied when the quasi-Fermi levels are flat across the junction ($V = V_\text{oc}$) and between the electrical contacts \cite{wurfel1982}. Further, Eqs. \ref{equation:Wurfel} and/or \ref{equation:Rau} may become invalid in the presence of non-equilibrium states such as radiative trap states \cite{zarrabi2020}, non-equilibrium excitons (as discussed above), and asymmetry in the light in and out-coupling due to optical interference or Stokes shift \cite{armin2020}. 

The radiative limit to the $V_\text{oc}$ can be calculated from detailed balance as \cite{rau2007}:
\begin{equation}
V_\text{oc}^\text{Rad} = \frac{k_\text{B}T}{q}\ln\left[\frac{J_\text{ph}}{J_0^\text{Rad}}+1\right]
\end{equation}
where $J_\text{ph}$ is the photocurrent generated at one-sun illumination ($J_\text{ph} = q\int_0^\infty \text{EQE}_\text{PV}\Phi_\text{sun}d E_\gamma$) and $J_0^\text{Rad}$ is the radiative dark saturation current ($J_0^\text{Rad} = q\int_0^\infty \text{EQE}_\text{PV}\Phi_\text{bb}d E_\gamma$). The true open-circuit voltage is the difference between the radiative limit and the total non-radiative losses ($\Delta V_\text{oc}^\text{NR,Total}$), defined as the sum of the intrinsic ($\Delta V_\text{oc}^\text{NR,Intrinsic}$) and electrode-induced ($\Delta V_\text{oc}^\text{NR,Electrode}$) losses: 
\begin{align}
qV_\text{oc} &= qV_\text{oc}^\text{Rad} - q\Delta V_\text{oc}^\text{NR,Total}\label{equation:qVoc}
\end{align}
The QFLS in the bulk of the active layer, ideally being equal to $\mu_\gamma$, is then determined by the difference between the radiative limit and the intrinsic non-radiative losses:
\begin{align}
\text{QFLS} &= qV_\text{oc}^\text{Rad} - q\Delta V_\text{oc}^\text{NR,Intrinsic}\label{equation:QFLS}
\end{align}
In general, the non-radiative losses of the $V_\text{oc}$ can be calculated from the external quantum efficiency $\eta_\text{LED}$ of a device operated as a light emitting diode (LED), defined as the ratio of the emitted radiation to injected current ($\eta_\text{LED} = q\Phi_\text{em}/J_\text{inj}$):
\begin{align}
q\Delta V_\text{oc}^\text{NR} &= -k_BT\ln\left[\eta_\text{LED}\right]\label{equation:deltaVocNR}
\end{align}
In order to isolate the effects of intrinsic and electrode-induced non-radiative loss one must consider the experimental conditions under which $\eta_\text{LED}$ is measured as well as the relation between $\Phi_\text{em}$ and Eqs.~\ref{equation:Wurfel} and \ref{equation:Rau}. Directly probing the emission described in Eq.~\ref{equation:Wurfel} would in the ideal case allow for the quantification of the intrinsic non-radiative losses to the $V_\text{oc}$ and subsequently the QFLS. 

\begin{figure*}
\includegraphics[scale=1.0]{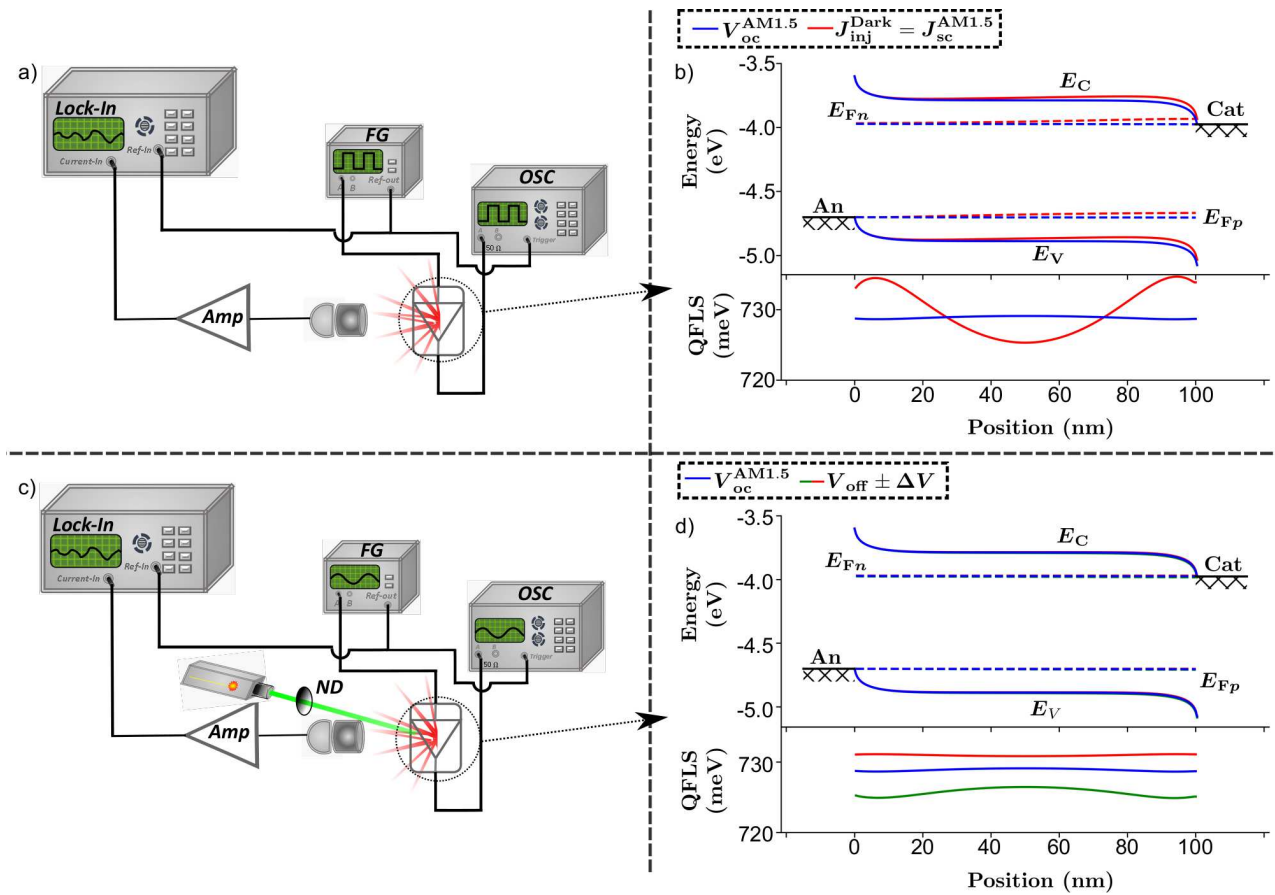}
\caption[]{(a) Electroluminescence schematic. (b) Simulated conduction ($E_\text{C}$) and valence levels ($E_\text{V}$) (solid lines), electron ($E_{\text{F}n}$) and hole ($E_{\text{F}p}$) quasi-Fermi levels (dashed lines), and QFLS under the conditions: one-sun open-circuit (blue) and dark injected current of one-sun short-circuit current (red).  (c) Electro-modulated photoluminescence schematic. (d) Simulated conduction, valence, electron and hole quasi-Fermi levels, and QFLS under the conditions: one-sun open-circuit (blue) and one-sun $V_\text{off}\pm\Delta V$ (red/green). An-anode, Cat-cathode, FG-function generator, Amp-current amplifier, OSC-oscilloscope, ND-neutral density wheel.}
\label{figure:expSetup}
\end{figure*}

\section{Experimental}

Figure \ref{figure:expSetup} shows two experimental apparatuses used to measure $\eta_\text{LED}$, the traditional electroluminescence (Figure \ref{figure:expSetup}(a)) and the alternative electro-modulated photoluminescence (Figure \ref{figure:expSetup}(c)). During electro\-luminescence measurements the device under test is held in the dark and a square-wave modulated voltage is applied to the device in forward bias by the function generator (Keysight 33500B). The current response can be measured as a square wave on the oscilloscope (Rohde \& Schwarz RTM3004) and the resulting electroluminescence is captured by a 50 mm lens, filtered with three low pass filters (Thorlabs FEL 600,550,550) (necessary to remove scattered pump light in the subsequent experiment), amplified by a silicon photoreceiver (FEMTO OE-300-Si-30), and measured with a lock-in amplifier (Stanford Research Systems SR860). The resulting external quantum efficiency ($\eta_\text{EL}$) can be calculated from the ratio of the electroluminescence intensity to the injected current at the reference frequency. This apparatus suffers from reduced geometric and spectral light collection efficiency which can be compensated for by calibrating the apparatus to an absolute measurement of $\eta_\text{EL}$ (see section S\Romannum{3}). To estimate $\eta_\text{EL}$ at a condition similar to open-circuit one-sun illumination, it is conventional to choose the injected dark current to equal the short-circuit current at one sun ($J_\text{inj}^\text{dark} = J_\text{sc}^\text{AM1.5}$). In contrast, Figure \ref{figure:expSetup}(c) shows the electro-modulated photoluminescence apparatus. Prior to performing an electro-modulated photoluminescence measurement the sample is illuminated with a laser (custom-made 520 nm diode laser) and the short circuit current ($J_\text{sc}$) is measured. To perform the electro-modulated photoluminescence quantum yield measurements the device is brought to open-circuit conditions where the applied time-dependent voltage, supplied by the function generator at angular frequency $\omega_V$, resulting injected current, measured by the oscilloscope, and emitted photoluminescence have the form:
\begin{align}
V_\text{app}(t) &= V_\text{off} + \Delta V\sin\left(\omega_V t\right)\label{equation:Vapp}\\
J_\text{inj}(t) &= \Delta J\sin\left(\omega_V t\right)\\
\Phi_\text{em}(t) &= \Phi_0 + \Delta \Phi\sin\left(\omega_V t\right)
\end{align}
where $\Delta J$ is kept smaller than $0.1J_\text{sc}$ by adjusting $\Delta V$ (see S S\Romannum{3} for details related to the size of this perturbation), and $V_\text{off}$ is set such that the mean current is zero. The resulting luminescence current amplitude ($\Delta\Phi$) is collected, amplified, and measured in the same manner as the electroluminescence measurement described above. The electro-modulated photoluminescence quantum efficiency is defined as the ratio between the luminescence intensity measured on the photodetector and the injected current amplitudes.
\begin{align}
\eta_\text{EMPL} = \frac{q\Delta \Phi}{\Delta J}
\end{align}
To evaluate $\eta_\text{EMPL}$ at conditions close to open-circuit one-sun illumination the laser power is adjusted such that the short-circuit current equals $J_\text{sc}^\text{AM1.5}$, leading $V_\text{off}$ to be approximately $V_\text{oc}^\text{AM1.5}$. As this experiment has the same spectral and geometric light collection efficiency as the electroluminescence measurement, the absolute $\eta_\text{EMPL}$ is found by multiplying by the same calibration factor.

\section{Results and Discussion}

Time-domain and steady-state drift-diffusion simulations (details in S\Romannum{1}) were employed to demonstrate the difference between electroluminescence and electro-modulated photoluminescence measurements in idealized systems before undertaking experiments. Figure \ref{figure:expSetup}(b) and (d) compare the simulated conduction, valence, electron and hole quasi-Fermi levels, and the QFLS across a device under one-sun open-circuit (blue), electroluminescence (Figure \ref{figure:expSetup}(b), red), and electro-modulated photoluminescence (Figure \ref{figure:expSetup} (d), red/green) conditions (see supplementary S\Romannum{2} for details about the electro-modulated photoluminescence simulations). Under electroluminescence conditions the conduction and valence levels, along with the quasi-Fermi levels, deviate from the one-sun open-circuit conditions as the QFLS varies across the device. In contrast, under the maximum and minimum voltages applied during electro-modulated photoluminescence the conduction and valence levels, and quasi-Fermi levels, are indistinguishable from those of one-sun open-circuit conditions. This indicates that electro-modulated photoluminescence conditions are closer to operational conditions than electroluminescence conditions. Crucially, under electro-modulated photoluminescence conditions the QFLS varies little across the device for each voltage and much less over the applied voltage range compared with the 10 meV variance in QFLS under $J_\text{inj}^\text{Dark} = J_\text{sc}^\text{AM1.5}$ conditions, seen in panel (b). The small variation in QFLS indicates that under electro-modulated photoluminescence conditions the device is very nearly in equilibrium with the emitted radiation, suggesting that the emission is described by Eq.~\ref{equation:Wurfel} and therefore can be used to quantify the QFLS. While the relatively large variation in QFLS under electroluminescence conditions indicates that the emission is dominated by the more general Eq.~\ref{equation:Rau}.

\begin{figure*}
\includegraphics[scale=1.0]{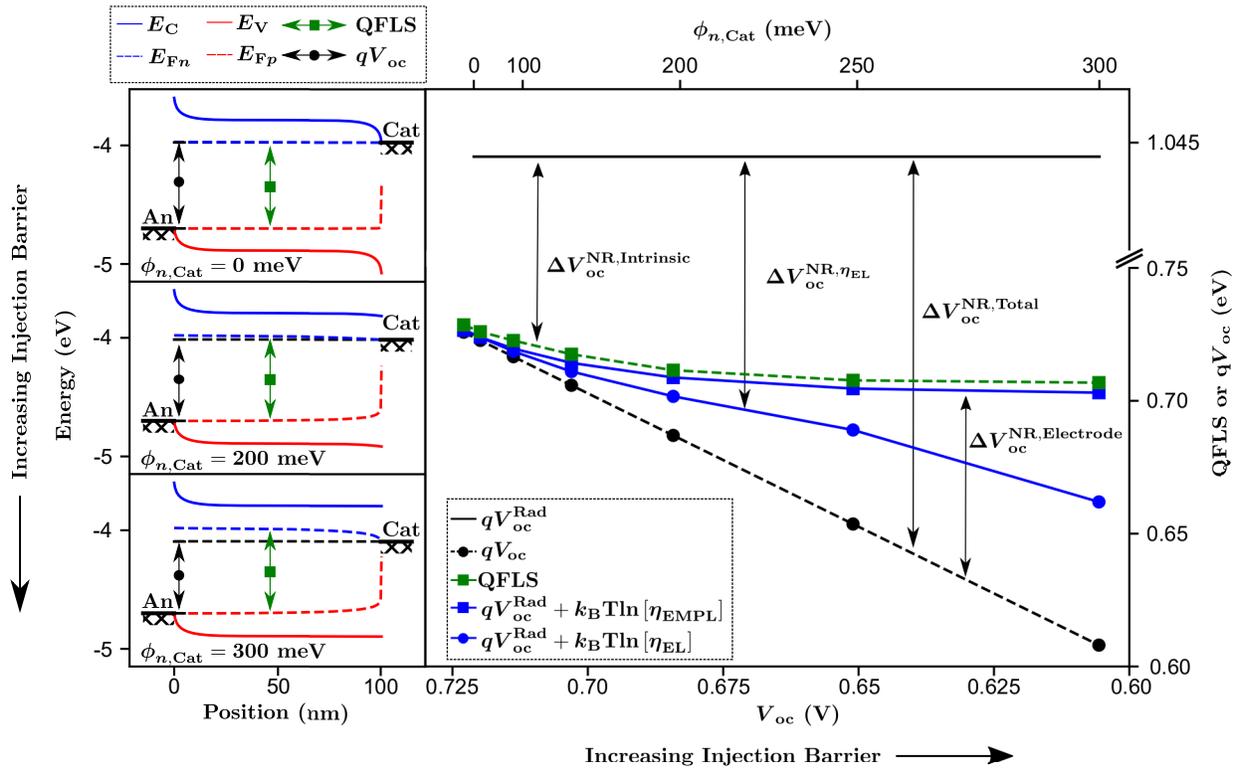}
\caption[]{Simulated (a-c) conduction ($E_\text{C}$), valence ($E_\text{V}$), electron ($E_{\text{F}n}$) and hole ($E_{\text{F}p}$) quasi-Fermi levels for devices with (a) no, (b) medium, and (c) high injection barrier. (d) Simulated injection barrier dependence of $V_\text{oc}$ (black circles), QFLS (green squares), and non-radiative losses measured by simulating electroluminescence (blue circles) and  electro-modulated photoluminescence (blue squares) experiments.}
\label{figure:SimulationvsPhi}
\end{figure*}

To explore the relationship between QFLS and $V_\text{oc}$ steady-state simulations of devices under open-circuit conditions were conducted. By increasing the electron injection barrier ($\phi_{n,\text{Cat}}$) at the cathode, the effect of an increased electrode-induced photovoltage loss can be simulated for organic photovoltaic devices \cite{sandberg2016,spies2017}. Figure \ref{figure:SimulationvsPhi}(a-c) show the energy levels for devices with no (a), small (b), and large (c) injection barriers. The QFLS is defined as the difference between the electron and hole quasi-Fermi levels in the bulk of the device (here taken to mean the exact center) as indicated by the green arrow, while the $V_\text{oc}$ can be determined by the difference between the electrode work functions as indicated by the black arrows \cite{sandberg2016}. As $\phi_{n,\text{Cat}}$ is increased the electron quasi-Fermi level near the cathode has to curve down in order to maintain equilibrium with the cathode work function. This leads to a considerable gradient in the electron quasi-Fermi level near the cathode ultimately reducing the $V_\text{oc}$ while the QFLS inside the bulk remains predominately unaffected. Figure \ref{figure:SimulationvsPhi}(d) summarizes these data for devices with electron injection barriers between 0 and 300 meV. In the absence of an injection barrier the QFLS and $qV_\text{oc}$ are identical, however, as the injection barrier is increased the $V_\text{oc}$ reduces while the QFLS is weakly affected. This is consistent with previous work where it was shown that $qV_\text{oc} \propto E_\text{G} - \phi_{n,\text{Cat}}$ for large enough electron injection barrier at the cathode \cite{sandberg2014,sandberg2016}. The size of the intrinsic and total non-radiative losses as well as those induced by the electrode ($\Delta V_\text{oc}^\text{NR,Electrode}$) and predicted by electroluminescence ($\Delta V_\text{oc}^{\text{NR},\eta_\text{EL}}$) are labelled for clarity. 

To uncover the relationship between the QFLS and $\eta_\text{LED}$, electroluminescence and electro-modulated photoluminescence experiments were simulated for each system. The non-radiative losses, calculated from Eq.~\ref{equation:deltaVocNR}, for both electroluminescence and electro-modulated photoluminescence are subtracted from the radiative limit; the blue curves in Figure \ref{figure:SimulationvsPhi} (d) show the results of these calculations. The predicted losses from electroluminescence, when subtracted from $V_\text{oc}^\text{Rad}$, follow $qV_\text{oc}$ for devices with low injection barrier, as expected from the reciprocity principle expressed in terms of Eq.~\ref{equation:Rau}. A discrepancy between the expected $qV_\text{oc}$ given by electroluminescence and the actual $qV_\text{oc}$ is seen for devices with large injection barrier. We note that an increased injection barrier at the cathode (reducing the number of injected electrons) generally makes the overall charge in the device uneven, which might explain the divergence from the expectations of the reciprocity principle. In contrast, the simulated losses obtained by electro-modulated photoluminescence follow the QFLS across all devices. This indicates that electro-modulated photoluminescence is probing the intrinsic losses occurring within the active layer of the device, useful for discerning material properties such as QFLS and optimizing electrodes in low-mobility systems, such as organic photovoltaic devices. While electro-luminescence is sensitive to both the intrinsic and electrode-induced non-radiative losses, useful in determining total non-radiative losses in an optimized device. Consequentially, based on this analysis, the open-circuit voltage expected from reciprocity $V_\text{oc,EL}$, and the QFLS in the bulk can be calculated from Eqs.~\ref{equation:qVoc}, \ref{equation:QFLS}, and \ref{equation:deltaVocNR} as:
\begin{align}
qV_\text{oc,EL} &= qV_\text{oc}^\text{Rad} + k_\text{B}T\ln\left[\eta_\text{EL}\right]\label{equation:qVocExpected} \text{,} \\
\text{QFLS} &= qV_\text{oc}^\text{Rad}+ k_\text{B}T\ln\left[\eta_\text{EMPL}\right]\label{equation:EQEEMQFLS} \text{.}
\end{align}

\begin{figure}
\includegraphics[scale=1.0]{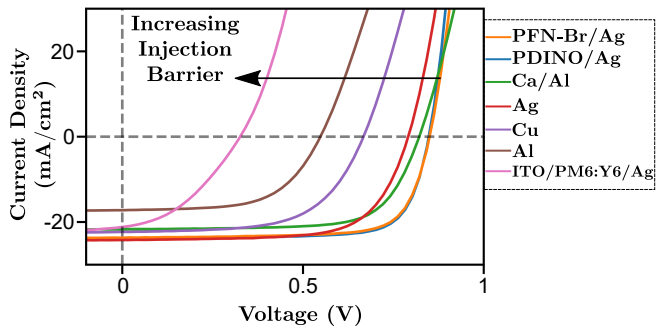}
\caption[]{Current density-voltage curves at one-sun illumination for devices with various cathode materials. All devices have an structure ITO/PEDOT:PSS/PM6:Y6/cathode (cathode material shown in legend), except for the pink curve where the device structure is ITO/PM6:Y6/Ag.}
\label{figure:expIV}
\end{figure}

\begin{figure*}
\includegraphics[scale=1.0]{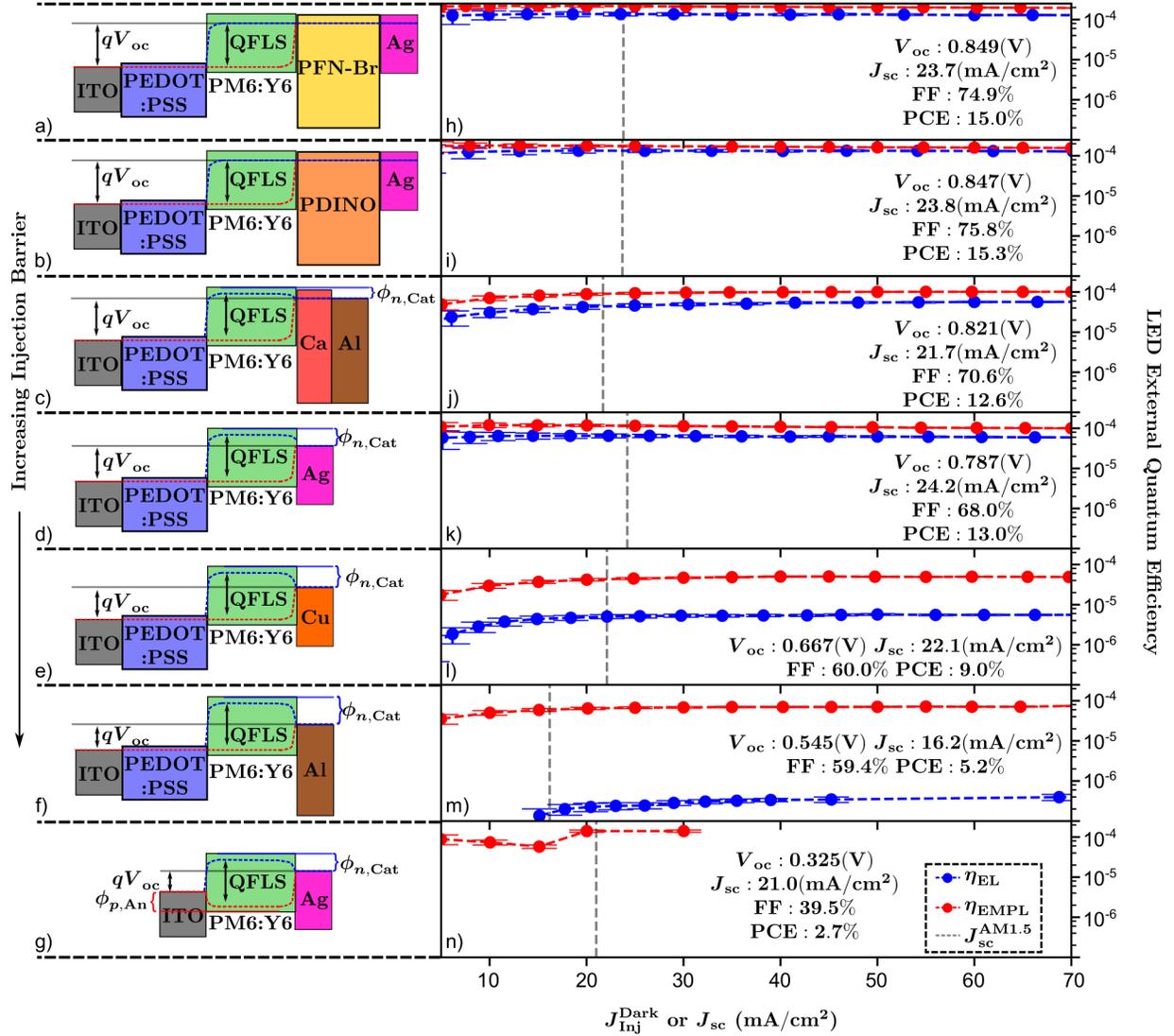}
\caption[]{(a-g) Device structures with increasing injection barrier. (h-n) LED external quantum efficiency as measured by electroluminescence ($\eta_\text{EL}$) as a function of dark injected current (blue) and electro-modulated photoluminescence ($\eta_\text{EMPL}$) as a function of short-circuit current (red), and $J_\text{sc}^\text{AM1.5}$ (grey dashed line). PCE-power conversion efficiency, FF-fill factor. Values of device parameters and measurements are taken from the top performing pixel. Error bars are calculated from measurement errors in the oscilloscope and lock-in amplifier (see S\Romannum{3} for details).}
\label{figure:EQEData}
\end{figure*}

To validate this proposal and the simulations (shown in Figure \ref{figure:SimulationvsPhi}) devices with cathode materials of different work functions were prepared to emulate the effect of varying the electron injection barrier at the cathode. Each device was made with ITO/PEDOT:PSS as the anode and a 100 nm PM6:Y6 active layer, while the cathode was varied to alter the electrode-induced non-radiative losses. Using this architecture an optimized device with power conversion efficiency of 15.3\% was created using PDINO/Ag as the cathode. Figure \ref{figure:expIV} shows the current-voltage characteristics for each device. Despite a relatively small shift in $J_\text{sc}$ and fill factor between the different devices the power conversion efficiency is reduced due to the reduction in $V_\text{oc}$, indicative of increasing electrode-induced non-radiative losses due to surface recombination (see supplemental Figure S3(a) and references \cite{sandberg2016,spies2017}). While the optimized device has a $V_\text{oc}$ of 0.847 V, the device using silver-only cathode exhibits a $V_\text{oc}$ of 0.787 V, the added interlayers (PDNIO or PFN-Br) modify the work function to create a more ohmic cathode. Figure \ref{figure:EQEData} (a-g) shows the device structures along with a sketch of the quasi-Fermi levels at open-circuit conditions. A near ohmic cathode (panels (a) and (b)) will cause the QFLS in the bulk and the $qV_\text{oc}$ to be roughly equivalent, due to the small injection barrier. Moving down from panel (a) to panel (f) the injection barrier is increased by increasing the cathode work function, causing a decrease in the $V_\text{oc}$ while the QFLS in the bulk of the active layer remains unchanged. The loss to the $V_\text{oc}$ is due to the increasing electrode-induced non-radiative loss associated with the increasing injection barrier. Figure \ref{figure:EQEData} (h-m) shows the LED external quantum efficiency measured by both electroluminescence and electro-modulated photoluminescence for each device as a function of dark injected current density or short-circuit current density respectively. As the injection barrier is increased $\eta_\text{EL}$ decreases until it is undetectable, while $\eta_\text{EMPL}$ does not substantially change. This suggests that $\eta_\text{EMPL}$ is measuring losses intrinsic to the semiconductor, while $\eta_\text{EL}$ is influenced by the structure of the device, as predicted by the simulations in Figure \ref{figure:SimulationvsPhi}. Figure \ref{figure:EQEData} (g) and (n) show the structure and external quantum efficiencies for a device without anode or cathode interlayers, severely limiting the driving force for charge extraction by creating an additional injection barrier at the anode. While  electroluminescence emission was not measurable in this case, the corresponding electro-modulated photoluminescence emission was. In this low $V_\text{oc}$ device $\eta_\text{EMPL}$ was similar to the other devices. This corroborates that electro-modulated photoluminescence measurements are insensitive to device structure and are measuring an intrinsic property of the active layer. 

\begin{figure*}
\includegraphics[scale=1.0]{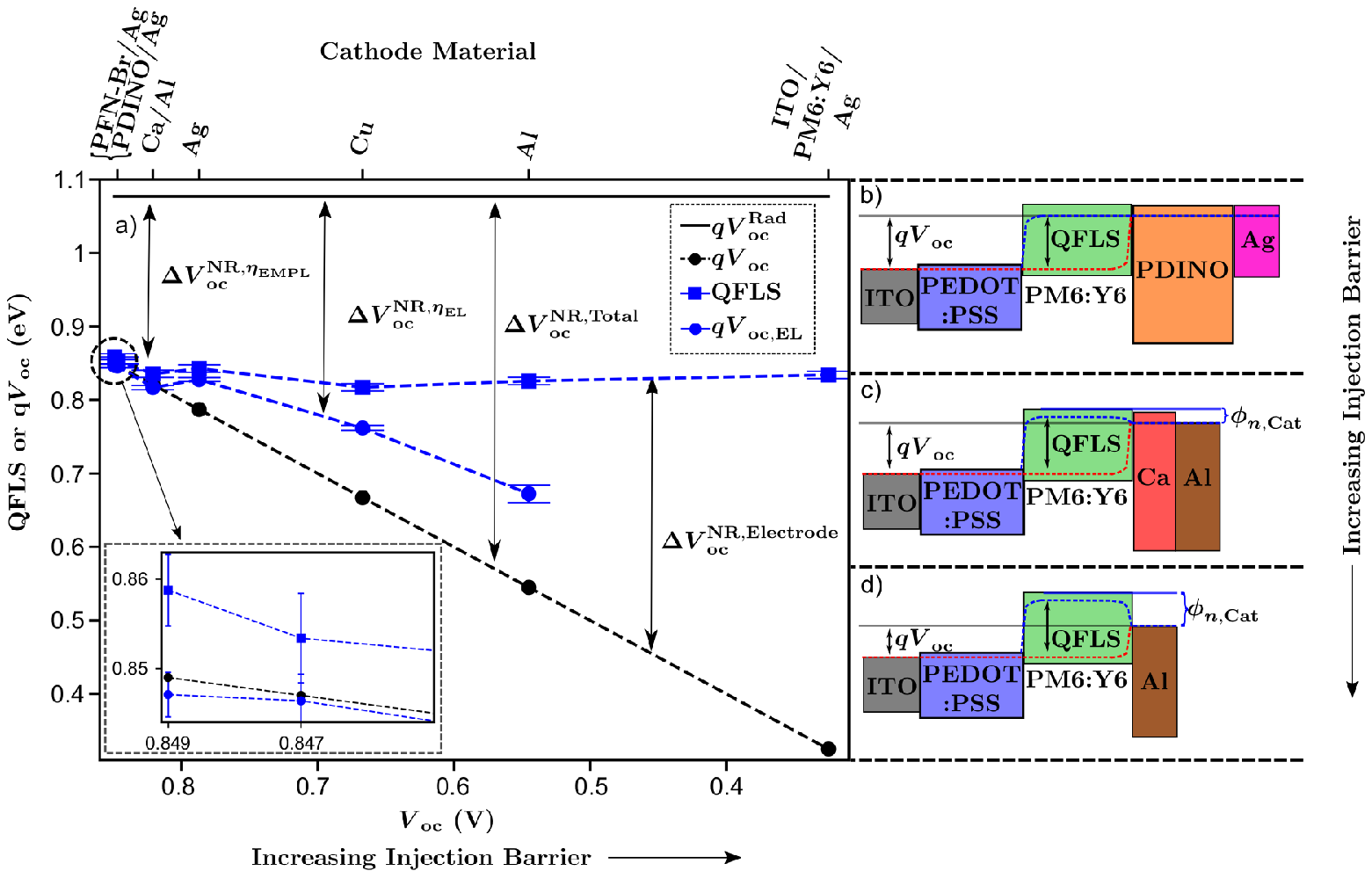}
\caption[]{(a) QFLS measured by electro-modulated photoluminescence (blue squares) and expected $qV_\text{oc}$ as measured by $\eta_\text{EL}$ (blue circles) for PM6:Y6 systems with increasing non-radiative $V_\text{oc}$ losses. Indicated on the plot is the size of the various non-radiative losses to the $V_\text{oc}$. Upper axis lists cathode materials while lower axis lists measured $V_\text{oc}^\text{AM1.5}$. (b-d) Device structure designed to have (b) low, (c) medium, and (d) high non-radiative losses at the cathode.}
\label{figure:ExpData}
\end{figure*}

The radiative limit to the $V_\text{oc}$ is calculated by measuring the photovoltaic external quantum efficiency on the optimized device and was found to be 1.077 V (see S\Romannum{3}). The expected $V_\text{oc}$ was calculated from $\eta_\text{EL}$ at $J_\text{Inj}^\text{Dark} = J_\text{sc}^\text{AM1.5}$ in accordance with Eq.~\ref{equation:qVocExpected}, and the QFLS is calculated from $\eta_\text{EMPL}$ at $J_\text{sc} = J_\text{sc}^\text{AM1.5}$ in accordance with Eq.~\ref{equation:EQEEMQFLS}. Figure \ref{figure:ExpData} (a) shows the QFLS, the expected $qV_\text{oc}$, and the measured $qV_\text{oc}$ at one-sun conditions for devices with increasing non-radiative losses. The black arrows indicate the magnitude of the various non-radiative losses as a guide for the eye. Figure \ref{figure:ExpData} (b-d) show examples of device structures with small (b), medium (c), and large (d)  injection barriers as a guide. The expected $qV_\text{oc}$ based on the measured electroluminescence follows the actual $qV_\text{oc}$ for devices with negligible injection barriers and occupy the region between the QFLS and $qV_\text{oc}$ for devices with impractical injection barriers, as predicted by the simulations in Figure \ref{figure:SimulationvsPhi}. The calculated QFLS is constant throughout all devices as expected, demonstrating the  precision of the method, and consistent with a previous study on the same material system using photo-induced absorption, further exemplifying the accuracy \cite{phuong2020}.

The inset of Figure \ref{figure:ExpData} (a) enhances the devices with lowest non-radiative losses. In the PDINO/Ag device ($V_\text{oc} = 0.847 $V) the QFLS is $853\pm 5$ meV, while the electrode-induced losses are not obvious due to the experimental uncertainty. However, in the PFN-Br device ($V_\text{oc} = 0.849$ V) the QFLS is $859\pm 4$ meV while the electrode-induced losses account for a reduction in the $V_\text{oc}$ of $10\pm 4$ meV. Improvements to the resolution will allow for discrepancies in the QFLS and the total non-radiative losses in an optimized cell to be distinguished, together quantifying the electrode-induced losses. This can be achieved by closely monitoring the temperature of the sample (accounting for about 3 meV of error) and decreasing the measurement error in the lock-in amplifier or the oscilloscope (accounting for about 1 meV of error each).

\section{Conclusions}
In conclusion, electro-modulated photoluminescence quantum yield was introduced as an approach to quantify the quasi-Fermi level splitting in the bulk of organic photovoltaic devices under operational conditions. Simulated electro-modulated photoluminescence measurements were compared to traditional electroluminescence measurements, used to quantify the total non-radiative losses. It was shown that electroluminescence predicted the total non-radiative loss when the cathode is sufficiently ohmic, and varies substantially when it is not, while electro-modulated photoluminescence accurately predicted the QFLS in the bulk of devices with both ohmic and non-ohmic cathodes. PM6:Y6 based devices were fabricated with various cathodes and both electro-modulated photoluminescence and electroluminescence measurements were performed. The experiments showed that with an increasingly non-ohmic cathode $\eta_\text{EL}$ decreased dramatically while $\eta_\text{EMPL}$ was unaffected, confirming the simulations. This showed that in the case of low-mobility systems, such as organic photovoltaic devices, $\eta_\text{EMPL}$ (along with the $V_\text{oc}$) fully characterizes different contributions of the non-radiative losses to the open-circuit voltage, providing a pathway to quantify the QFLS. It was found that in a high $V_\text{oc}$ optimized solar cell the QFLS was $859\pm 4$ meV, leading to the conclusion that the electrode-induced losses accounted for $10\pm 4$ mV of the total photovoltage loss. This approach was the first direct measurement of QFLS within organic solar cells and will contribute to the continual improvement of solar cell and light-emitting diode performance by allowing one to distinguish between intrinsic and electrode-induced losses to the open-circuit voltage.

\begin{acknowledgments}
This work was supported by the S\^er Cymru \Romannum{2} through the European Regional Development Fund, Welsh European Funding Office, and Swansea University strategic initiative in Sustainable Advanced Materials. A.A. is a S\^er Cymru II Rising Star Fellow and P.M. is a S\^er Cymru II National Research Chair. This work was also funded by UKRI through the EPSRC Program Grant EP/T028511/1 Application Targeted Integrated Photovoltaics. Drew B. Riley acknowledges the support of the Natural Sciences and Engineering Research Council of Canada (NSERC), [PGSD3-545694-2020]. Ronald \"Osterbacka and Nora M. Wilson would like to acknowledge support from Svenska tekniska vetenskapsakademien i Finland and The Society of Swedish Literature in Finland.
\end{acknowledgments}

\clearpage
\bibliography{QuantificationOfQFLS}

\end{document}